\documentclass[10pt]{amsart}
\usepackage{amsbsy,amssymb,amscd,amsfonts,latexsym,amstext,delarray,
amsmath,amsthm,graphicx}

\input xypic

\newtheorem{thm}{Theorem}[section]

\newtheorem{cor}[thm]{Corollary}
\newtheorem{lem}[thm]{Lemma}

\newtheorem{defn}[thm]{Definition}

\newtheorem{ex}[thm]{Example}

\numberwithin{equation}{section}

\def\bL{{\mathbb L}}

\def\A{{\mathbb A}}
\def\C{{\mathbb C}}
\def\F{{\mathbb F}}

\renewcommand{\P}{{\mathbb P}}
\def\Q{{\mathbb Q}}
\def\Z{{\mathbb Z}}
\def\R{{\mathbb R}}

\def\m{{\mathfrak{m}}}

\def\cancel#1#2{\ooalign{$\hfil#1\mkern1mu/\hfil$\crcr$#1#2$}}

\def\cutp{{\rm p}}
\def\cutq{{\rm q}}
\def\cuts{{\rm s}}
\def\cutu{{\rm u}}
\def\cutv{{\rm v}}

\def\cA{{\mathcal A}}
\def\cB{{\mathcal B}}

\def\cE{{\mathcal E}}

\def\cI{{\mathcal I}}

\def\cL{{\mathcal L}}
\def\cM{{\mathcal M}}
\def\cN{{\mathcal N}}
\def\cO{{\mathcal O}}

\def\cQ{{\mathcal Q}}
\def\cR{{\mathcal R}}
\def\cS{{\mathcal S}}

\def\cU{{\mathcal U}}
\def\cV{{\mathcal V}}
\def\cX{{\mathcal X}}
\def\cY{{\mathcal Y}}

\newcommand{\ie}{{\it i.e.\/}\ }
\newcommand{\eg}{{\it e.g.\/}\ }
\newcommand{\cf}{{\it cf.\/}\ }
\def\text{\hbox}

\def\Ker{{\rm Ker}}

\title{Supermanifolds from Feynman graphs}
\author[Marcolli]{Matilde Marcolli}
\author[Rej]{Abhijnan Rej}
\address{Max--Planck Institut f\"ur Mathematik  \\
Vivatsgasse 7 \\
Bonn, D-53111 Germany} \email{marcolli\@@mpim-bonn.mpg.de}
\email{rej\@@mpim-bonn.mpg.de}

\begin{document}

\maketitle

\begin{abstract}
We generalize the computation of Feynman integrals of log divergent graphs in terms of the 
Kirchhoff polynomial to the case of graphs with both fermionic and bosonic edges, to which we 
assign a set of ordinary and Grassmann variables. This procedure gives a computation of the 
Feynman integrals in terms of a period on a supermanifold, for graphs admitting a basis of 
the first homology satisfying a condition generalizing the log divergence in this context. The
analog in this setting of the graph hypersurfaces is a graph supermanifold given by the divisor of zeros and poles of the Berezinian of a matrix associated to the graph, inside a superprojective space. We introduce a Grothendieck group for supermanifolds and we identify the subgroup generated by the graph supermanifolds. This can be seen as a general procedure to construct interesting classes of supermanifolds with associated periods.
\end{abstract}

\tableofcontents

\section{Introduction}

The investigation of the relation between Feynman integrals and
motives originates in the work of Broadhurst and Kreimer
\cite{BroKr}, where it is shown that zeta and multiple zeta values
appear systematically in the evaluation of Feynman diagrams. These
are very special periods, namely they are believed to arise as
periods of mixed Tate motives. An
important question in the field then became understanding the a priori
reason for the appearance of this special class of motives in
quantum field theory. Surprisingly, the work of Belkale and Brosnan
\cite{BeBro} revealed a universality result for the varieties
associated to Feynman graphs, namely they generate the Grothendieck
ring of varieties. This means that, as motives, they
can be arbitrarily far from the mixed Tate case. The question then
moved on to whether the piece of the cohomology of the graph
hypersurface complement, which is involved in the evaluation of the
Feynman integral as a period, actually happens to be mixed Tate. The
recent results of \cite{BEK}, see also \cite{Blo}, analyze this
problem in depth in the case of the ``wheels with n-spokes'' graphs.
There are considerable technical difficulties involved in the
cohomological calculations, even for relatively uncomplicated
graphs, due to the singularities of the graph hypersurfaces and to
the fact that generally their complexity grows very rapidly with
the combinatorial complexity of the graphs. A different approach to
the relation between Feynman integrals and mixed Tate motives was
given by Connes--Marcolli in \cite{CoMa}, from the point of view of
Tannakian categories and motivic Galois groups. This approach originated
from the earlier work of Connes--Kreimer \cite{CoKr} where it is
shown that the Feynman graphs of a given physical theory form a
commutative, non-cocommutative, Hopf algebras. This defines 
dually an affine group scheme, called the group of diffeographisms 
of the theory, whose Lie algebra bracket is given by the symmetrized 
insertion of one graph into another at vertices. 
The Connes--Kreimer Hopf algebra structure of perturbative
renormalization was extended from the case of scalar field theories
to the case of QED, and more general gauge theories, by van
Suijlekom in \cite{WvS}. He showed that the Ward identities define a
Hopf ideal in the Connes--Kreimer Hopf algebra of Feynman graphs.
A related question of motivic lifts of the Connes--Kreimer Hopf algebra 
is formulated in \cite{Blo}.  

\smallskip

The fact that the graph hypersurfaces generate the Grothendieck ring of varieties 
means that the computation of the Feynman integral in terms of a period on the
complement of a graph hypersurface in a projective space gives a general procedure 
to construct a large class of interesting varieties with associated periods. Our purpose 
here is to show that this general procedure can be adapted to produce a large
class of interesting supermanifolds with associated periods.

In the setting of \cite{BEK} and \cite{Blo} one is assuming, from
the physical viewpoint, that all edges of the graph are of the same
nature, as would be the case in a scalar field theory with
Lagrangian
\begin{equation}\label{Lagr}
 \cL(\phi)= \frac 12 (\partial \phi)^2 - \frac{m^2}{2} \phi^2 -
\cL_{int}(\phi).
\end{equation}
However, in more general theories, one has graphs that are
constructed out of different types of edges, which correspond to
different propagators in the corresponding Feynman rules. We
consider the case of theories with fermions, where graphs
have both {\em fermionic} and {\em bosonic} legs. From the
mathematical point of view, it is natural to replace the usual
construction of the graph hypersurface by a different construction
which assigns to the edges either ordinary variables (bosonic)
or Grassmann variables (fermionic). This procedure yields a
natural way to construct a family of {\em supermanifolds} associated
to this type of Feynman graphs. 

We give a computation of the Feynman integral in terms of a bosonic
and a fermionic integration, so that the integral is computed as a
period on a supermanifold that is the complement of a divisor in 
a superprojective space, defined by the set of zeros and poles of the Berezinian
of a matrix $\cM(t)$ associated to a graph $\Gamma$ and a choice of a basis
$B$ for $H_1(\Gamma)$. We refer to the divisor defined by this Berezinian as
the graph supermanifold $\cX_{(\Gamma,B)}$.

As in the case of the ordinary graph hypersurfaces, we are interested in 
understanding their motiving nature first by looking at their classes in 
the Grothendieck ring of varieties. To this purpose, we introduce a
Grothendieck ring $K_0(\cS\cV_\C)$ of supermanifold and we prove that it is a polynomial 
ring $K_0(\cV_\C)[T]$ over the Grothendieck ring of ordinary varieties. We then use this
result to prove that the classes of the graph supermanifolds $\cX_{(\Gamma,B)}$
generate the subring $K_0(\cV_\C)[T^2]$, where the degree two appears due to
a fermion doubling used in the computation of the Feynman integral.

In a different perspective, an interest in supermanifolds and their
periods has recently surfaced in the context of mirror symmetry, see 
\cite{Se}, \cite{AgVa}, \cite{KuPo}. We do not know, at present, whether the
classes of supermanifolds considered here and their periods may be
of any relevance to that context. We mention some points of contact
in \S \ref{MSsect} below.

As the referee pointed out to us, a theory of parametric Feynman
integrals for scalar supersymmetric theories was developed in
\cite{Medv}. The type of integrals we are considering here is slightly
different from those of \cite{Medv}, hence we cannot apply directly
the results of that paper. It would be interesting to see what
class of graph supermanifolds can be obtained from the
parametric integrals of \cite{Medv}. 

\medskip

{\bf Acknowledgment.} The first author is partially 
supported by NSF grant DMS-0651925.
The second author is supported as a Marie Curie Early Stage Researcher
at Durham University and by the Clay Mathematical Institute.

\medskip

\subsection{Graph varieties and periods}

The evaluation of Feynman integrals in perturbative quantum field
theory can be expressed, in the case of logarithmically divergent
graphs (which have $n$ loops and $2n$ edges), in terms of a period
in the algebro--geometric sense. This is obtained as the integration
over a simplex of an algebraic differential form involving the graph
polynomial of the Feynman graph (\cf \cite{Blo}, \cite{BEK})
\begin{equation}\label{Fint}
\int_\sigma \frac{\Omega}{\Psi_\Gamma^2},
\end{equation}
with $\Psi_\Gamma$ the graph polynomial (Kirchhoff polynomial)
of the graph $\Gamma$, $\sigma$ the simplex in $\P^{2n-1}$, and
\begin{equation}\label{Omegasum}
 \Omega=\sum_{i=1}^{2n} (-1)^i x_i\, dx_1 \cdots \widehat{dx_i}\cdots
dx_n.  
\end{equation}
The logarithmically divergent case is the one where periods
are defined independently of a renormalization procedure. In the
more general case, the problem arises from the fact that the
integrand acquires poles along exceptional divisors in the blowup
along faces of the simplex (see \cite{BEK}, \cite{Blo}).

In the following, given a graph $\Gamma$ we denote by $\Psi_\Gamma$
the graph polynomial
\begin{equation}\label{PsiGamma}
\Psi_\Gamma(x) =\sum_{T \subset \Gamma} \prod_{e\notin T} x_e,
\end{equation}
where the sum is over all the spanning trees $T$ of $\Gamma$ and the
product is over edges not belonging to $T$. These give homogeneous
polynomials in the variables $x=(x_e)=(x_1,\cdots,x_{\# E(\Gamma)})$
associated to the edges of $\Gamma$, where each variable appears of
degree at most one in each monomial. They define hypersurfaces
\begin{equation}\label{XGamma}
X_\Gamma =\{ x=(x_e)\in \P^{\# E(\Gamma)-1}\, |\, \Psi_\Gamma (x)=0 \}.
\end{equation}
These are typically singular hypersurfaces.

In the case of the log divergent graphs considered in \cite{BEK},
the motive involved in the evaluation of the Feynman integral as a
period is of the form
$$ H^{2n-1}(P\smallsetminus Y_\Gamma, \Sigma \smallsetminus (\Sigma \cap
Y_\Gamma)), $$ where $n$ is the number of loops, $P \to \P^{2n-1}$
is a blowup along linear spaces, $Y_\Gamma$ is the strict transform
of $X_\Gamma$ and $\Sigma$ is the total inverse image of the
coordinate simplex of $\P^{2n-1}$.

The recent results of Bergbauer--Rej
\cite{BeRe} provide an explicit combinatorial formula for the graph
polynomial under insertion of one graph into another.

\subsection{Grothendieck ring}

Recall that the Grothendieck ring $K_0(\cV_K)$ of varieties is
generated by quasi-projective varieties over a field $K$ with the
relation
\begin{equation}\label{relGrV}
[X] = [Y] + [X\smallsetminus Y],
\end{equation}
for $Y\subset X$ a closed subvariety. It is made into a ring by the
product of varieties.

Based on computer experiments, which showed that many graph
hypersurfaces satisfy the condition
$$ \# X_\Gamma(\F_q) = P_\Gamma (q), $$
for some polynomial $P_\Gamma$, Kontsevich conjectured that the
$X_\Gamma$ would be always mixed Tate. The main result of
Belkale--Brosnan \cite{BeBro} disproved the conjecture by showing
that the classes $[X_\Gamma]$ are very general. In fact, they span
the Grothendieck ring of varieties, which means that the $X_\Gamma$
can be quite arbitrary as motives. As discussed in \cite{BEK} and \cite{Blo}, it is
especially interesting to construct explicit stratifications of the
graph hypersurfaces and try to identify which strata are likely to
be non-mixed-Tate.

In the Grothendieck ring $K_0(\cV_\C)$ the class $[\A^1]=[\C]$ is often denoted by $\bL$ 
and is the class of the Lefschetz motive, with $[\P^1]= 1+\bL$ and $1=[pt]=[\A^0]$. 

There are two opposite ways to deal with the Lefschetz motive $\bL$. If, as in the theory of motives, one formally inverts $\bL$, one enriches in this way the Grothendieck ring of varieties by the Tate motives $\bL^n$, $n\in \Z$. In the theory of motives, one usually denotes $\Q(1)$ the formal inverse of the Lefschetz motive, with $\Q(n)=\Q(1)^{\otimes n}$.  The category of pure (respectively, mixed) Tate motives is the subcategory of the abelian (respectively, triangulated) category of motives generated by the $\Q(n)$.

If, instead, one maps the Lefschetz motive $\bL$ to zero, one obtains the
semigroup ring of stable birational equivalence classes of varieties, by the result of \cite{LaLu}, 
which we briefly recall. Two irreducible varieties $X$ and $Y$ are said to be
stably birationally equivalent if $X\times \P^n$ is birational to
$Y\times \P^m$ for some $n,m\geq 0$.
It is proved in \cite{LaLu}, that there is a ring isomorphism
\begin{equation}\label{GrVSB}
K_0[\cV_\C]/I \cong \Z[SB],
\end{equation}
where $SB$ is the semigroup of stable birational classes of
varieties with the product induced by the product of varieties,
$\Z[SB]$ is the associated semigroup ring, and $I\subset
K_0[\cV_\C]$ is the ideal generated by the class $[\A^1]$ of the
affine line.
The result of \cite{LaLu} essentially depends on the
Abramovich--Karu--Matsuki--Wlodarczyk factorization theorem
\cite{AKMW}, which shows that any rational birational map of smooth
complete varieties decomposes as a sequence of blowups and
blowdowns, and on Hironaka's resolution of singularities.

\section{Supermanifolds and motives}

\subsection{Supermanifolds}

We recall here a few basic facts of supergeometry that we need in
the following. The standard reference for the theory of
supermanifolds is Manin's \cite{Man}.

By a complex supermanifold one understands a datum $\cX=(X,\cA)$ with
the following properties: $\cA$ is a sheaf of supercommutative
rings on $X$; $(X,\cO_X)$ is a complex manifold, where
$\cO_X=\cA/\cN$, with $\cN$ the ideal of nilpotents in
$\cA$; the quotient $\cE=\cN/\cN^2$ is locally free over
$\cO_X$ and $\cA$ is locally isomorphic to the exterior algebra
$\Lambda^\bullet_{\cO_X}(\cE)$, where the grading is the
$\Z_2$-grading by odd/even degrees. The supermanifold is split if
the isomorphism $\cA \cong \Lambda^\bullet_{\cO_X}(\cE)$ is
global.

\begin{ex} Projective superspace. {\rm The complex projective
superspace $\P^{n|m}$ is the supermanifold $(X,\cA)$ with 
$X=\P^n$ the usual complex projective space and 
$$ \cA =\Lambda^\bullet ( \C^m \otimes_\C \cO(-1) ), $$
with the exterior powers $\Lambda^\bullet$ graded by odd/even degree.
It is a split supermanifold.}
\end{ex}

A {\em morphism} $F: \cX_1 \to \cX_2$ of supermanifolds
$\cX_i=(X_i,\cA_i)$, $i=1,2$, consists of a pair $F=(f,f^\#)$ of a 
morphism of the underlying 
complex manifolds $f: X_1 \to X_2$ together with a morphism 
$f^\#:\cA_2 \to f_*\cA_1$ of sheaves of supercommutative rings
with the property that at each point $x\in X_1$ the induced
morphism $f^\#_x: (\cA_2)_{f(x)} \to (\cA_1)_x$ satisfies
$f^\#_x(\m_{f(x)})\subset \m_x$, on the maximal ideals of 
germs of sections vanishing at the point (\cf \cite{Man}, \S 4.1).

\smallskip

In particular, an {\em embedding} of complex supermanifolds is a
morphism $F=(f,f^\#)$ as above, with the property that $f:X_1 
\hookrightarrow X_2$ is an embedding and $f^\#:\cA_2 \to f_*\cA_1$ 
is surjective. As in ordinary geometry, we define the ideal sheaf of
$\cX_1$ to be the kernel
\begin{equation}\label{sheafId}
\cI_{\cX_1}:= \Ker( f^\#: \cA_2 \to f_*\cA_1 ).
\end{equation}

\smallskip

An equivalent characterization of an embedding of supermanifold is
given as follows. If we denote by $E_i$, for $i=1,2$ the holomorphic
vector bundles on $X_i$ such that $\cO(E_i)=\cE_i=\cN_i/\cN_i^2$, with
the notation as above, then an embedding  $F: \cX_1 \hookrightarrow
\cX_2$ is an embedding $f:X_1 \hookrightarrow
X_2$ such that the induced morphism of vector bundles $f^*: E_2 \to E_1$ is 
surjective (\cf \cite{LePooWe}).
Thus, we say that $\cY=(Y,\cB)$ is a closed sub-supermanifold of
$\cX=(X,\cA)$ when there exists a closed embedding $Y \subset X$ and
the pullback of $E_\cA$ under this embedding surjects to $E_\cB$.

\smallskip

An open submanifold $\cU=(U,\cB) \hookrightarrow \cX=(X,\cA)$ is given 
by an open embedding $U\hookrightarrow X$ of the underlying complex 
manifolds and an isomorphism of sheaves $\cA |_U \cong \cB$. When $\cY\subset \cX$
is a closed embedding and $U=X\smallsetminus Y$, the ideal sheaf of
$\cY$ satisfies $\cI_\cY |_U =\cA |_U$.

\smallskip

A subvariety in superprojective space is a supermanifold 
\begin{equation}\label{projsub}
\cX=(X\subset \P^n, (\Lambda^\bullet(\C^m\otimes_\C \cO(-1))/\cI)|_X),
\end{equation}
where $\cI=\cI_\cX$ is an ideal generated by finitely many homogeneous 
polynomials of given $\Z/2$-parity. In other words, if we denote by
$(x_0,\ldots,x_n,\theta_1,\ldots,\theta_m)$ the bosonic and fermionic
coordinates of $\P^{n|m}$, then a projective subvariety can be
obtained by assigning a number of equations of the form
\begin{equation}\label{poleqsuper}
\Psi^{ev/odd}(x_0,\ldots,x_n,\theta_1,\ldots,\theta_m)=
\sum_{i_1< \cdots <i_k} P_{i_1,\ldots,i_k}(x_0,\ldots,x_n)
\theta_{i_1}\cdots \theta_{i_k}=0,
\end{equation}
where the $P_{i_1,\ldots,i_k}(x_0,\ldots,x_n)$ are homogeneous
polynomials in the bosonic variables. 

\smallskip

Notice that there are strong constraints in supergeometry on realizing
supermanifolds as submanifolds of superprojective space. For instance,
Penkov and Skornyakov \cite{PenSko} showed that super Grassmannians 
in general do not embed in superprojective space, \cf \cite{Man}.
The result of LeBrun, Poon, and Wells \cite{LePooWe} shows that a
supermanifold $\cX=(X,\cA)$ with compact $X$ can be embedded in some
superprojective space $\P^{n|m}$ if and only if it has a positive
rank-one sheaf of $\cA$-modules. 

\smallskip

Notice that, in the above, we have been working with complex
projective superspace and complex subvarieties. However, it is
possible to consider supergeometry in an arithmetic context,
as shown in \cite{SchwSha}, so that it makes sense to investigate
extensions of motivic notions to the supergeometry setting.
In the present paper we limit our investigation of motivic aspects 
of supermanifolds to the analysis of their classes in a suitable
Grothendieck ring.

\subsection{A Grothendieck group}

We begin by discussing the Grothendieck group of varieties in the 
supergeometry context and its relation to the Grothendieck group of
ordinary varieties.

We first recall the following notation from \cite{KaSh} \S II.2.3.
Given a locally closed subset $Y\subset X$ and a sheaf $\cA$ on $X$,
there exists a unique sheaf $\cA_Y$ with the property that
\begin{equation}\label{Arestr}
\cA_Y|_Y=\cA|_Y \ \ \text{ and } \ \  \cA_Y|_{X\smallsetminus Y}=0. 
\end{equation}
In the case where $Y$ is closed, this satisfies $\cA_Y =i_*(\cA|_Y)$ where
$i:Y\hookrightarrow X$ is the inclusion, and when $Y$ is open it
satisfies $\cA_Y=\Ker(\cA\to i_*(\cA|_{X\smallsetminus Y}))$.

\begin{defn}\label{SGrV}
Let $\cS\cV_\C$ be the category of complex supermanifolds with
morphisms defined as above. Let $K_0(\cS\cV_\C)$ denote the free
abelian group generated by the isomorphism classes of objects $\cX\in
\cS\cV_\C$ subject to the following relations. 
Let $F:\cY \hookrightarrow \cX$ be a closed
embedding of supermanifolds. Then 
\begin{equation}\label{relSGr}
[\cX]=[\cY]+[\cX\smallsetminus \cY],
\end{equation}
where $\cX\smallsetminus \cY$ is the supermanifold 
\begin{equation}\label{Ucomplement}
\cX\smallsetminus \cY = (X\smallsetminus Y, \cA_X |_{X\smallsetminus Y}).
\end{equation}
\end{defn}

In particular, in the case where $\cA=\cO_X$ is the 
structure sheaf of $X$, the relation \eqref{relSGr} reduces to the usual
relation 
\begin{equation}\label{GrVrel} 
[X]=[Y]+[X \smallsetminus Y]. 
\end{equation}
in the Grothendieck group of ordinary varieties, for a closed
embedding $Y\subset X$.

\begin{lem}\label{splitGr}
All supermanifolds decompose in $K_0(\cS\cV_\C)$ as a finite
combination of split supermanifolds, and in fact of supermanifolds
where the vector bundle $E$ with $\cO(E)=\cE=\cN/\cN^2$ is trivial.
\end{lem}

\proof This is a consequence of the d\'evissage of coherent sheaves. 
Namely, for any coherent sheaf $\cA$ over a Noetherian reduced 
irreducible scheme there exists a dense open set $U$ such that 
such that $\cA|_U$ is free. 
The relation \eqref{relSGr} then ensures that, given a pair
$\cX=(X,\cA)$ and the sequence of sheaves
$$ 0 \to i_{!}(\cA|_U)\to \cA \to j_*(\cA|_Y) \to 0, $$
associated to the open embedding $U\subset X$ with complement
$Y=X\smallsetminus U$, the class $[X,\cA]$ satisfies
$$ [X,\cA] = [U,\cA_U|_U] +[Y,\cA_Y|_Y]. $$
The sheaf $\cA_Y$ on $X$, which has support $Y$, has a chain of
subsheaves $\cA_Y\supset \cA_1\supset\cdots\supset\cA_k=0$ such that
each quotient $\cA_i/\cA_{i+1}$ is coherent on $Y$. Thus, one can
find a stratification where on each open stratum the supermanifold
is split and with trivial vector bundle. The supermanifold
$\cX=(X,\cA)$ decomposes as a sum of the corresponding classes
in the Grothendieck group.
\endproof

The fact that the vector bundle that constitutes the fermionic part of
a supermanifold is trivial when seen in the Grothendieck group is
the analog for supermanifolds of the fact that any projective bundle is
equivalent to a product in the Grothendieck group of ordinary varieties.

\smallskip

It follows from Lemma \ref{splitGr} above that the product makes
$K_0(\cS\cV_\C)$ into a ring with 
$$ [\cX][\cY]=[\cX\times \cY]. $$ 
In fact, we have the following more precise description of
$K_0(\cS\cV_\C)$ in terms of the Grothendieck ring of ordinary varieties.

\begin{cor}\label{ringSGr}
The Grothendieck ring $K_0(\cS\cV_\C)$ of supervarieties is a
polynomial ring over the Grothendieck ring of ordinary varieties of
the form
\begin{equation}\label{SGrPoly}
K_0(\cS\cV_\C)=K_0(\cV_\C)[T],
\end{equation}
where $T=[\A^{0|1}]$ is the class of the affine superspace of
dimension $(0,1)$.
\end{cor}

It then follows that the relation \eqref{GrVSB} between the
Grothendieck ring and the semigroup ring of stable birational
equivalence classes extends to this context.

\smallskip

Notice that, in the supermanifold case, there are now two different types of Lefschetz motives, namely the bosonic one $\bL_b=[\A^{1|0}]$ and the fermionic one $\bL_f=[\A^{0|1}]$. By analogy to what happens in motivic integration and in the theory of motives, we may want to localize at the Lefschetz motives, \ie invert both $\bL_b$ and $\bL_f$. That is, according to Corollary \ref{ringSGr}, we consider the field of fractions of $K_0(\cV_\C)[\bL_b^{-1}]= \cS^{-1} K_0(\cV_\C)$, with respect to the multiplicative semigroup $\cS=\{ 1, \bL_b, \bL_b^2,\ldots \}$ and then the ring of Laurent polynomials 
\begin{equation}\label{TateSGr}
 \cS^{-1}K_0(\cV_\C)[\bL_f, \bL_f^{-1}]=K_0(\cV_\C)[\bL_b^{-1}, \bL_f, \bL_f^{-1}]. 
\end{equation}
This suggests extensions of motivic integration to the context of supermanifolds, but we will not pursue this line of thought further in the present paper.

\smallskip

There is also a natural extension to supermanifolds of the usual notion of
birational equivalence. We say that two supermanifolds $\cX=(X,\cA)$
and $(Y,\cB)$ are birational if there exist open dense embeddings of
supermanifolds $\cU\subset \cX$ and $\cV\subset \cY$ and
an isomorphism $\cU \cong \cV$. Similarly, one can give a notion
analogous to that of stable birational equivalence by requiring that
there are superprojective spaces $\P^{n|m}$ and $\P^{r|s}$ such that
$\cX\times \P^{n|m}$ and $\cY\times \P^{r|s}$ are birational. One then
finds the following. We denote by $\Z[SSB]$ the semigroup ring of
stable birational equivalence classes of supermanifolds.

\begin{cor}\label{SGrSSB}
There is a surjective ring homomorphism $K_0(\cS\cV_\C)\to \Z[SSB]$,
which induces an isomorphism
\begin{equation}\label{SGrVSB}
K_0(\cS\cV_\C)/I \cong \Z[SSB],
\end{equation}
where $I$ is the ideal generated by the classes $[\A^{1|0}]$ and
$[\A^{0|1}]$. 
\end{cor}

The formal inverses of $\bL_f$ and $\bL_b$ correspond to two types of Tate objects 
for motives of supermanifold, respectively fermionic and bosonic Tate motives.
We see from Corollary \ref{ringSGr} and \eqref{TateSGr} that the fermionic part 
of the supermanifolds only contribution to the class in the Grothendieck ring is
always of this fermonic Tate type, while only the bosonic part can provide non-Tate 
contributions.

\subsection{Integration on supermanifolds}

The analog of the determinant in supergeometry is given by the Berezinian. This is defined in
the following way. Suppose given a matrix $\cM$ of the form
$$ \cM =\left(\begin{array}{cc} M_{11} & M_{12} \\ M_{21} & M_{22} \end{array}\right), $$
where the $M_{11}$ and $M_{22}$ are square matrices with entries of order zero and the $M_{12}$ and $M_{21}$ have elements of order one. Then the Berezinian of $\cM$ is the expression
\begin{equation}\label{BerM}
{\rm Ber} (\cM):= \frac{\det(M_{11} - M_{12} M_{22}^{-1} M_{21})} {\det(M_{22})}. 
\end{equation}
It satisfies ${\rm Ber}(\cM\cN)={\rm Ber}(\cM){\rm Ber}(\cN)$.

It is shown in \cite{BeLei} that Grassmann integration satisfies a change of variable formula where the Jacobian of the coordinate change is given by the Berezinian ${\rm Ber}(J)$ with $J$ the matrix
$J=\frac{\partial X_\alpha}{\partial Y_\beta}$ with $X_\alpha=(x_i,\xi_r)$ and $Y_\beta=(y_j,\eta_s)$.
We explain in \S \ref{SupGraphSec} below how to use this to replace expressions of the form
\eqref{Fint} for Feynman integrals, with similar expressions involving a Berezinian and a Grassmann integration over a supermanifold.

\subsection{Divisors}

There is a well developed theory of divisors on supermanifolds,
originating from \cite{RSV}. A Cartier divisor on a supermanifold $\cX$ of
dimension $(n|m)$ is defined by a collection of {\em even} meromorphic
functions $\phi_i$ defined on an open covering $\cU_i\hookrightarrow
\cX$, with $\phi_i \phi_j^{-1}$ a holomorphic function on $\cU_i \cap
\cU_j$ nowhere vanishing on the underlying $U_i\cap U_j$. Classes of
divisors correspond to equivalence classes of line bundles and can be
described in terms of integer linear combinations of
$(n-1|m)$-dimensional subvarieties $\cY \subset \cX$. 

\section{Supermanifolds from graphs}\label{SupGraphSec}

\subsection{Feynman's trick and Schwinger parameters}\label{FtrickSec}

We begin by describing a simple generalization of the well known ``Feynman
trick'', 
$$ \frac{1}{ab} = \int_0^1 \frac{1}{(xa+(1-x)b)^2} dx, $$
which will be useful in the following.  The results recalled here are well 
known in the physics literature (see \eg \cite{BjDr} \S 8 and
\S 18), but we give a brief and self contained treatment here for the 
reader's convenience. A similar derivation from a more 
algebro-geometric viewpoint can be found in \cite{BEK}.

\begin{lem}\label{FeynmanTrick}
Let $\Sigma_n$ denote the $n$-dimensional simplex 
\begin{equation}\label{Sigman}
\Sigma_n=\{ (t_1,\ldots,t_n)\in (\R^*_+)^n \,|\, \sum_{i=1}^n t_i \leq
1 \} .
\end{equation}
Let $dv_{\Sigma_n}=dt_1\cdots dt_{n-1}$ be the volume form on $\Sigma_n$ induced by the
standard Euclidean metric in $\R^n$.
Then, for given nonzero parameters $q_i$, for $i=1,\ldots,n$, the following identity holds:
\begin{equation}\label{Trick}
\frac{1}{q_1\ldots q_n} = (n-1)! \, \int_{\Sigma_{n-1}} \frac{1}{(t_1 q_1 + \cdots +
t_n q_n)^n} dv_{\Sigma_n},
\end{equation}
where $t_n =1-\sum_{i=1}^{n-1} t_i$.
\end{lem}

\proof The following identity holds:
\begin{equation}\label{prodqkGamma}
\frac{1}{q_1^{k_1}\cdots q_n^{k_n}} = \frac{1}{\Gamma(k_1) \cdots \Gamma(k_n)} 
\int_0^\infty \cdots \int_0^\infty e^{-(s_1 q_1 +\cdots + s_n q_n)}\, s_1^{k_1-1} \cdots s_n ^{k_n-1}
 \,\,ds_1 \cdots ds_n.
\end{equation}
The $s_i$ are usually called Schwinger parameters in the physics literature. We then perform a change of variables, by setting $S=\sum_{i=1}^n s_i$ and $s_i = S t_i$, with $t_i\in [0,1]$ with $\sum_{i=1}^n t_i =1$. Thus, we rewrite \eqref{prodqkGamma} in the form
\begin{equation}\label{prodqkSt}
\frac{1}{q_1^{k_1}\cdots q_n^{k_n}} = \frac{\Gamma(k_1+\cdots+k_n)}{\Gamma(k_1) \cdots \Gamma(k_n)}  \int_0^1 \cdots \int_0^1 \frac{t_1^{k_1-1} \cdots t_n^{k_n-1}\, \delta(1-\sum_{i=1}^n t_i)}{(t_1 q_1 +\cdots+ t_n q_n)^{k_1+\cdots +k_n}} 
 \,\,dt_1 \cdots dt_n.
\end{equation}
The result \eqref{Trick} then follows as a particular case of this more general identity, with $k_i=1$ for $i=1,\ldots,n$ and $\Gamma(n)=(n-1)!$.
\endproof

One can also give an inductive proof of \eqref{Trick} by Stokes theorem, which avoids introducing any transcendental functions, but the argument we recalled here is standard and it suffices for our purposes.

\smallskip

The Feynman trick is then related to the graph polynomial
$\Psi_\Gamma$ in the following way (see again \cite{BjDr}, \S 18 
and \cite{Naka}).
Suppose given a graph $\Gamma$. Let $n=\# E(\Gamma)$ be the number of
edges of $\Gamma$ and let $\ell=b_1(\Gamma)$ be the number of loops,
\ie the rank of $H^1(\Gamma,\Z)$. Suppose chosen a set of generators 
$\{ l_1,\ldots, l_\ell \}$ of $H^1(\Gamma,\Z)$. We then define the
$n\times \ell$-matrix $\eta_{ik}$ as
\begin{equation}\label{etaik}
\eta_{ik}=\left\{ \begin{array}{rl} +1 & \text{edge $e_i\in$ loop
$l_k$, same orientation} \\[2mm] -1 & \text{edge $e_i\in$ loop
$l_k$, reverse orientation} \\[2mm] 0 & \text{otherwise.} \end{array}\right.
\end{equation}
Also let $M_\Gamma$ be the $\ell\times \ell$ real symmetric matrix 
\begin{equation}\label{MGamma}
(M_\Gamma)_{kr}(t)=\sum_{i=0}^n t_i \eta_{ik} \eta_{ir}, 
\end{equation}
for $t=(t_0,\ldots,t_{n-1})\in \Sigma_n$ and $t_n=1-\sum_i t_i$.
Let $s_k$, $k=1,\ldots,\ell$ be real variables $s_k\in \R^D$ assigned
to the chosen basis of the homology $H^1(\Gamma,\Z)$. Also let $p_i$,
for $i=1,\ldots,n$ be real variables $p_i\in \R^D$ associated to the
edges of $\Gamma$. Let $q_i(p)$ denote the quadratic form
\begin{equation}\label{qp}
q_i(p)=p_i^2 - m_i^2,
\end{equation}
for fixed parameters $m_i > 0$. These correspond to the Feynman
propagators 
\begin{equation}\label{bosonpropag}
 \frac{1}{q_i} = \frac{1}{p_i^2 - m_i^2} 
\end{equation}
for a scalar field theory, associated by the Feynman rules to the
edges of the graph. One can make a change of variables
$$ p_i = u_i + \sum_{k=1}^\ell \eta_{ik} s_k, \ \ \text{ with the constraint } \ \ 
 \sum_{i=0}^n t_i u_i \eta_{ik} =0. $$
Then we have the following result.

\begin{lem}\label{intMGamma}
The following identity holds
\begin{equation}\label{intDsk}
\int \frac{1}{(\sum_{i=0}^n t_i q_i)^n} \,\, d^Ds_1\cdots
d^Ds_\ell = C_{\ell,n} \, \det(M_\Gamma(t))^{-D/2} (\sum_{i=0}^n t_i
(u_i^2 - m_i^2))^{-n+D\ell/2}.
\end{equation}
\end{lem}

\proof 
After the change of variables, the left hand side reads
$$ \int \frac{d^Ds_1\cdots d^Ds_\ell}
{(\sum_{i=0}^n t_i (u_i^2 - m_i^2) + \sum_{kr}
(M_\Gamma)_{kr} s_k s_r)^n} \,\,  . $$
The integral can then be reduced by a change of variables that
diagonalizes the matrix $M_\Gamma$ to an integral of the form
$$ \int \frac{d^Dx_1\cdots d^Dx_\ell}{(a-\sum_k \lambda_k x_k^2)^n} =
C_{\ell,n} \, a^{-n+D\ell/2} \prod_{k=1}^\ell \lambda_k^{-D/2} , $$
with
$$ C_{\ell,n} = \int \frac{d^Dx_1\cdots d^Dx_\ell}{(1-\sum_k x_k^2)^n}. $$ 
\endproof

This is the basis for the well known formula that relates the
computation of Feynman integrals to periods, used in \cite{BEK}. 
In fact, we have the following.

\begin{cor}\label{intandperiod}
In the case of graphs where the number of
edges and the number of loops are related by $n=D\ell/2$, 
the Feynman integral is computed by
\begin{equation}\label{FeynPer}
\int \frac{d^Ds_1\cdots d^Ds_\ell}{q_0 \cdots q_n} = C_{\ell,n}\,
\int_{\Sigma_n} \frac{dt_0\cdots
dt_{n-1}}{\Psi_\Gamma(t_0,\ldots,t_n)^{D/2}},
\end{equation}
where $t_n=1-\sum_{i=0}^{n-1}
t_i$ and
\begin{equation}\label{MPsiGamma}
\Psi_\Gamma(t)=\det(M_\Gamma(t)).
\end{equation}
 \end{cor} 

\proof
Notice that, in the case of graphs with $n=D\ell/2$, the
integration \eqref{intDsk} reduces to
\begin{equation}\label{intDsk0}
 \int \frac{d^Ds_1\cdots d^Ds_\ell}
{(\sum_{i=0}^n t_i q_i)^n}  =C_{\ell,n}\, \det(M_\Gamma(t))^{-D/2}. 
\end{equation}
\endproof

We now consider a modified version of this construction, where we deal
with graphs that have both bosonic and fermionic legs, and we maintain
the distinction between these two types at all stages by
assigning to them different sets of ordinary and Grassmann
variables. Strictly from the physicists point of view this is an
unnecessary complication, because the formulae we recalled in this
section adapt to compute Feynman integrals also in theories with
fermionic fields, but from the mathematical viewpoint this procedure
will provide us with a natural way of constructing an interesting 
class of supermanifolds with associated periods. 

\subsection{The case of Grassmann variables}

Consider now the case of Feynman propagators and Feynman diagrams that
come from theories with both bosonic and fermionic fields. This means
that, in addition to terms of the form \eqref{Lagr}, the Lagrangian
also contains fermion interaction terms. The form of such terms is
severely constrained (see \eg \cite{Ramond}, \S 5.3): for instance, in
dimension $D=4$ renormalizable interaction terms can only involve at
most two fermion and one boson field. 

The perturbative expansion for such theories
corresponsingly involve graphs $\Gamma$ with two different types of
edges: fermionic and bosonic edges. The Feynman rules assign to each
bosonic edge a propagator of the form \eqref{bosonpropag} and to
fermionic edges a propagator
\begin{equation}\label{fermipropag}
i\frac{\cutp + m}{p^2 - m^2} = \frac{i}{\cutp -m}.
\end{equation}
Notice that in physically significant theories one would have $i({\mathpalette\cancel p} -m)^{-1}$
with ${\mathpalette\cancel p}=p^\mu\gamma_\mu$, but for simplicity we work here with
propagators of the form \eqref{fermipropag}, without tensor indices.

In the following we use the notation
\begin{equation}\label{qcutq}
q(p)=p^2 - m^2, \ \ \ \ \cutq(p)=i(\cutp + m)
\end{equation}
for the quadratic and linear forms that appear in the propagators
\eqref{bosonpropag} and \eqref{fermipropag}.  In the following, again just to 
simplify notation,  we also drop the mass terms in the propagator (\ie we set $m=0$)
and ignore the resulting infrared divergence problem. The reader can easily reinstate
the masses whenever needed.

Thus, the terms of the form $(q_1
\cdots q_n)^{-1}$, which we encountered in the purely bosonic case, 
are now replaced by terms of the form
\begin{equation}\label{fermiprodqi}
\frac{\cutq_1 \cdots \cutq_f}{q_1 \cdots q_n},
\end{equation}
where $n=\# E(\Gamma)$ is the total number of edges in the graph and 
$f=\# E_f(\Gamma)$ is the number of fermionic edges.

\medskip

We first prove an analog of Lemma \ref{FeynmanTrick}, where we now
introduce an extra set of Grassmann variables associated to the 
fermionic edges. The derivation we present suffers from a kind of
``fermion doubling problem'', in as each fermionic edge contributes
an ordinary integration variables, which essentially
account for the denominator $q_i$ in \eqref{fermipropag} and
\eqref{fermiprodqi}, as well as a {\em pair} of Grassman variables 
accounting for the numerator $\cutq_i$ in \eqref{fermipropag} 
and \eqref{fermiprodqi}. 

Let $\cQ_f$ denote the $2f\times 2f$ antisymmetric 
matrix
\begin{equation}\label{cQf}
\cQ_f=\left( \begin{array}{ccccccc}  
0 & \cutq_1 & 0 & 0 & \cdots & 0 & 0 \\
-\cutq_1 & 0 & 0 & 0 & \cdots & 0 & 0 \\
0 & 0 & 0 & \cutq_2 & \cdots & 0 & 0 \\
0 & 0 & -\cutq_2 & 0 & \cdots & 0 & 0 \\
\vdots & & & & \cdots & & \vdots \\
0 & 0 & 0 & 0 & \cdots & 0 & \cutq_f \\
0 & 0 & 0 & 0 & \cdots & -\cutq_f & 0 
\end{array} \right).
\end{equation}

\begin{lem}\label{SigmanfInt}
Let $\Sigma_{n|2f}$ denote the superspace $\Sigma_n \times
\A^{0|2f}$. 
Then the following identity holds:
\begin{equation}\label{qcutqSigmanf}
\frac{\cutq_1 \cdots \cutq_f}{q_1 \cdots q_n} = K_{n,f}
\int_{\Sigma_{n|2f}} \frac{dt_1\cdots dt_{n-1} d\theta_1
\cdots d\theta_{2f}}{(t_1 q_1 +\cdots t_n q_n +
\frac{1}{2}\theta^t \cQ_f \theta)^{n-f}}  ,
\end{equation}
with 
$$ K_{n,f}=\frac{2^f (n-1)!}{\prod_{k=1}^f (-n+f-k+1)}. $$
\end{lem}

\proof We first show that the following identity holds for integration in the
Grassmann variables $\theta=(\theta_1,\ldots,\theta_{2f})$:
\begin{equation}\label{intQtheta}
 \int \frac{d\theta_1 \cdots d\theta_{2f}}{(1+\frac{1}{2}\theta^t
\cQ_f \theta)^\alpha} =
\frac{f!}{2^f} {{-\alpha}\choose{f}} \cutq_1 \cdots \cutq_f.
\end{equation}
In fact, we expand using the Taylor series
$$ (1+x)^\beta = \sum_{k=0}^\infty {{\beta}\choose{k}} x^k $$
and the identity
$$ \frac{1}{2}\theta^t \cQ_f \theta= \sum_{i=1}^f \cutq_i
\theta_{2i-1}\theta_{2i}, $$
together with the fact that the degree zero variables
$x_i=\theta_{2i-1}\theta_{2i}$ commute, to obtain
$$ (1+\frac{1}{2}\theta^t \cQ_f \theta)^{-\alpha} =
\sum_{k=0}^\infty {{-\alpha}\choose{k}} ( \sum_{i=1}^f \cutq_i
\theta_{2i-1}\theta_{2i})^k . $$
The rules of Grassmann integration then imply that only the 
coefficient of $\theta_1\cdots \theta_{2f}$ remains as a result of the
integration. This gives \eqref{intQtheta}. 

For simplicity of notation, we then write $T=t_1 q_1 +\cdots t_n
q_n$, so that we have
$$ \int_{\Sigma_{n|2f}} \frac{1}{(t_1 q_1 +\cdots t_n q_n + 
\frac{1}{2}\theta^t \cQ_f \theta)^{n-f}}  dt_1\cdots dt_{n-1}\, d\theta_1
\cdots d\theta_{2f} = $$ $$  \frac{f!}{2^f} {{-n+f}\choose{f}}
\cutq_1 \cdots \cutq_f
\int_{\Sigma_n} T^{-n+f} T^{-f} \,\,dt_1\cdots dt_{n-1} $$   
$$ = \frac{f!}{2^f} {{-n+f}\choose{f}} \cutq_1 \cdots \cutq_f
\int_{\Sigma_n} \frac{dt_1\cdots dt_{n-1}}{(t_1 q_1 +\cdots +t_n
q_n)^n} = \frac{f!}{2^f (n-1)!} {{-n+f}\choose{f}}
\frac{\cutq_1 \cdots \cutq_f}{q_1\cdots q_n}. 
$$
\endproof

\subsection{Graphs with fermionic legs}

Consider now the case of graphs that have both bosonic and fermionic
legs. We mimic the procedure described above, but by using both
ordinary and Grassmann variables in the process. 

We divide the edge indices $i=1,\ldots,n$ into two sets
$i_b=1,\ldots,n_b$ and $i_f=1,\ldots,n_f$, with $n=n_b+n_f$,
respectively labeling the bosonic and fermionic legs. Consequently,
given a choice of a basis for the first homology of the graph, indexed
as above by $r=1,\ldots,\ell$, we replace the matrix $\eta_{ir}$ of
\eqref{etaik}, with a matrix of the form 
\begin{equation}\label{etabf}
\left( \begin{array}{cc} \eta_{i_f r_f} & \eta_{i_f r_b}  \\ \eta_{i_b r_f} & \eta_{i_b r_b}  
\end{array} \right).
\end{equation}
Here the loop indices $r=1,\ldots,\ell$ are at first divided into
three sets $\{ 1,\ldots,\ell_{ff} \}$, labelling the loops consisiting
of only fermionic edges, $\{ 1,\ldots, \ell_{bb}\}$ labelling the
loops consisting of only bosonic edges, and the remaning variables $\{
1,\ldots,\ell_{bf} =\ell-(\ell_{ff}+\ell_{bb})\}$ for the loops that
contain both fermionic and bosonic edges.  We then introduce two sets
of momentum variables: ordinary variables $\cuts_{r_b}\in \A^{D|0}$,
with $r_b=1,\ldots, \ell_b=\ell_{bb}+\ell_{bf}$, and Grassmann
variables $\sigma_{r_f}\in \A^{0|D}$ with $r_f=1,\ldots,
\ell_f=\ell_{ff}+\ell_{bf}$. That is, we assign to each purely
fermionic loop a Grassmann momentum variable, to each purely bosonic
loop an ordinary momentum variable, and to the loops containing both
fermionic and bosonic legs a pair $(\cuts_r,\sigma_r)$ of an ordinary
and a Grassman variable.  In \eqref{etabf} above we write $r_f$ and
$r_b$, respectively, for the indexing sets of these Grassmann and
ordinary variables.   

We then consider a change of variables 
\begin{equation}\label{varchange}
\cutp_{i_b} = \cutu_{i_b} + \sum_{r_f} \eta_{i_b r_f} \sigma_{r_f} +
\sum_{r_b} \eta_{i_b r_b} \cuts_{r_b}, \ \ \ \  \cutp_{i_f}
=\cutu_{i_f} + \sum_{r_f} \eta_{i_f r_f} \sigma_{r_f} + \sum_{r_b}
\eta_{i_f r_b} \cuts_{r_b}. 
\end{equation} 
analogous to the one used before, where now, for reasons of
homogeneity, we need to assume that the $\eta_{i r_f}$ are of degree
one and the $\eta_{i r_b}$ are of degree zero, since the $\cutp_i$ are
even (ordinary) variables. 

We apply the change of variables \eqref{varchange} to the expression
\begin{equation}\label{vartochange}
\sum_i t_i p_i^2 + \sum_{i_f}  \theta_{2i_f-1}\theta_{2i_f} \cutp_{i_f}.
\end{equation}

We assume again, as in the purely bosonic case (\cf (18.35) of \cite{BjDr}), the relations
$$ \sum_i t_i \cutu_i \eta_{i r} =0 $$
for each loop variable $r=r_b$ and $r=r_f$.

We can then rewrite \eqref{vartochange} in the form
$$ \sum_i t_i u_i^2 + \sum_{i_f}  \theta_{2i_f-1}\theta_{2i_f}  \cutu_{i_f} $$
$$  + \sum_{r_b, r_b'} (\sum_i t_i \eta_{i r_b} \eta_{i r_b'}) \cuts_{r_b} \cuts_{r_b'} -
\sum_{r_f r_f'} (\sum_i t_i \eta_{i r_f} \eta_{i r_f'}) \sigma_{r_f} \sigma_{r_f'}  $$
$$ + \sum_{r_b r_f} \left((\sum_i t_i \eta_{i r_b} \eta_{i r_f}) \cuts_{r_b} \sigma_{r_f}-
\sigma_{r_f}^\tau \cuts_{r_b}^\tau (\sum_i t_i \eta_{i r_f} \eta_{i r_b})\right) $$
$$  + \sum_{r_b} (\sum_{i_f}  \theta_{2i_f-1}\theta_{2i_f} \eta_{i_f r_b}) \cuts_{r_b} +
\sum_{r_f} (\sum_{i_f}  \theta_{2i_f-1}\theta_{2i_f} \eta_{i_f r_f}) \sigma_{r_f} .$$
Notice the minus sign in front of the quadratic term in the $\sigma_{r_f}$, since 
for order-one variables $ \sigma_{r_f} \eta_{ir_f'} = -  \eta_{i r_f'} \sigma_{r_f} $.
We write the above in the simpler notation
\begin{equation}\label{shortform}
 T + \cuts^\tau M_b(t) \cuts - \sigma^\tau M_f(t) \sigma+ \sigma^\tau M_{fb}(t) \cuts - \cuts^\tau M_{bf}(t) \sigma  + N_b(\theta) \cuts + N_f(\theta) \sigma, 
\end{equation}
where $\tau$ denotes transposition, $\cuts=(\cuts_{r_b})$, $\sigma=(\sigma_{r_f})$, and
\begin{equation}\label{shortnotation}
\begin{array}{l}
T=\sum_i t_i u_i^2 + \sum_{i_f}  \theta_{2i_f-1}\theta_{2i_f}  \cutu_{i_f}, \\[2mm] 
M_b(t)=\sum_i t_i \eta_{i r_b} \eta_{i r_b'}, \\[2mm] 
M_f(t)=\sum_i t_i \eta_{i r_f} \eta_{i r_f'}= - M_f(t)^\tau, \\[2mm]
M_{fb}(t)=\sum_i t_i \eta_{i r_b} \eta_{i r_f}, \\[2mm]
N_b(\theta)=\sum_{i_f}  \theta_{2i_f-1}\theta_{2i_f} \eta_{i_f r_b}, \\[2mm]
N_f(\theta)=\sum_{i_f}  \theta_{2i_f-1}\theta_{2i_f} \eta_{i_f r_f}.
\end{array}
\end{equation}
Since the $\eta_{i,r_f}$ are of degree one and the $\eta_{i,r_b}$ of degree zero, the matrices $M_b$ and $M_f$ are of degree zero,  the $M_{bf}$ and $M_{fb}$ of degree one, while the $N_b$ and $N_f$ are, respectively, of degree zero and one.  Thus, the expression \eqref{shortform} is of degree zero. Notice that, since the $\eta_{i r_f}$ are of order one, the matrix $M_f(t)$ is antisymmetric. We also set
$M_{bf}(t)=M_{fb}(t)=M_{fb}(t)^\tau$.

\medskip

We then consider an integral of the form
$$ \int \frac{d^D\cuts_1 \cdots d^D\cuts_{\ell_b}\, d^D\sigma_1 \cdots d^D\sigma_{\ell_f}}
{(\sum_i t_i p_i^2 + \sum_{i_f}  \theta_{2i_f-1}\theta_{2i_f} \cutp_{i_f})^{n-f}}  =  $$
\begin{equation}\label{SupInt}
 \int \frac{d^D\cuts_1 \cdots d^D\cuts_{\ell_b}\, d^D\sigma_1 \cdots d^D\sigma_{\ell_f}}
{(T + \cuts^\tau M_b(t) \cuts + N_b(\theta) \cuts -\sigma^\tau M_f(t) \sigma + \sigma^\tau M_{fb}(t) \cuts - \cuts^\tau M_{fb}(t)^\tau \sigma + N_f(\theta) \sigma)^{n-f}},
\end{equation}
where the $d^D\sigma_i=d\sigma_{i1}\cdots d\sigma_{iD}$ are Grassmann variables integrations and the $d^D \cuts_i$ are ordinary integrations.

\medskip

Recall that for Grassmann variables we have the following change of variable formula.

\begin{lem}\label{Grchvar}
Suppose given an invertible antisymmetric $N\times N$-matrix $A$ with entries of degree zero and an $N$-vector $J$ with entries of degree one. Then we have
\begin{equation}\label{sigmaetachange}
\sigma^\tau A \sigma + \frac{1}{2} (J^\tau \sigma-\sigma^\tau J) =\eta^\tau A \eta 
+ \frac{1}{4}  J^\tau A^{-1} J,
\end{equation}
for $\eta= \sigma - \frac{1}{2}A^{-1}J$.
\end{lem}

\proof The result is immediate: since $A^\tau= -A$, we simply have
$$ \eta^\tau A \eta = \sigma^\tau A \sigma + \frac{1}{2} J^\tau \sigma - \frac{1}{2} \sigma^\tau J - \frac{1}{4} J^\tau A^{-1} J. $$
\endproof

We then use this change of variable to write
\begin{equation}\label{quadsigmaeta}
\begin{array}{c}
-\sigma^\tau M_f(t) \sigma + \sigma^\tau M_{fb}(t) \cuts - \cuts^\tau M_{fb}(t)^\tau \sigma + \frac{1}{2} (
\sigma^\tau N_f(\theta) -N_f(\theta)^\tau \sigma) = \\[3mm]
-\eta^\tau M_f(t) \eta - \frac{1}{4} (M_{fb}(t)\cuts + \frac{1}{2} N_f(\theta))^\tau M_f(t)^{-1}  (M_{fb}(t)\cuts + \frac{1}{2} N_f(\theta)) \end{array}
\end{equation}
with
\begin{equation}\label{etasigma}
\eta= \sigma -\frac{1}{2} M_f(t)^{-1} \left(M_{fb}(t) \cuts + \frac{1}{2} N_f(\theta) \right).
\end{equation}

We have
$$  \frac{1}{4} (M_{fb}(t)\cuts + \frac{1}{2} N_f(\theta))^\tau M_f(t)^{-1}  (M_{fb}(t)\cuts + \frac{1}{2} N_f(\theta)) = $$
$$ \frac{1}{4}  \cuts^\tau M_{bf}(t) M_f(t)^{-1} M_{fb}(t) \cuts + \frac{1}{8} (
 N_f(\theta)^\tau M_f(t)^{-1} M_{fb}(t)\cuts + \cuts^\tau M_{bf}(t) M_f(t)^{-1} N_f(\theta) ) $$ $$
 + \frac{1}{16} N_f(\theta)^\tau M_f(t)^{-1} N_f(\theta). $$

We then let 
\begin{equation}\label{Utthetas}
U(t,\theta,\cuts):= T+ C(t,\theta) +\cuts^\tau A_b(t) \cuts + B_b(t,\theta) \cuts,
\end{equation}
where 
\begin{equation}\label{AbtBbttheta}
\begin{array}{l}
A_b(t)= M_b(t)-\frac{1}{4} M_{bf}(t) M_f(t)^{-1} M_{fb}(t) \\[3mm]
B_b(t,\theta)=N_b(\theta)-\frac{1}{4} N_f(\theta)^\tau M_f(t)^{-1} M_{fb}(t) \\[3mm]
C(t,\theta)=-\frac{1}{16} N_f(\theta)^\tau M_f(t)^{-1} N_f(\theta).
\end{array}
\end{equation}

Thus, we write the denominator of \eqref{SupInt} in the form
\begin{equation}\label{denomprod}
U(t,\theta,\cuts)^{n-f} \left( 1+ \frac{1}{2} \eta^\tau X_f(t,\theta,\cuts) \eta \right)^{n-f},
\end{equation}
where we use the notation
\begin{equation}\label{XfMfU}
X_f(t,\theta,\cuts):= 2 U(t,\theta,\cuts)^{-1} M_f(t).
\end{equation}

Thus, the Grassmann integration in \eqref{SupInt} gives, as in Lemma \ref{SigmanfInt},
\begin{equation}\label{inteta}
\int \frac{d^D\eta_1 \cdots d^D \eta_{\ell_f}}{ \left( 1+ \frac{1}{2} \eta^\tau X_f(t,\theta,\cuts) \eta \right)^{n-f}}  = C_{n,f,\ell_f}\, \frac{2^{D\ell_f/2}}{U(t,\theta,\cuts)^{D\ell_f/2}} \det(M_f(t))^{D/2},
\end{equation}
where $C_{n,f,\ell_f}$ is a combinatorial factor obtained as in Lemma \ref{SigmanfInt}.

\medskip

We then proceed to the remaining ordinary integration in \eqref{SupInt}. We have, dropping a multiplicative constant,
\begin{equation}\label{intspart}
\det(M_f(t))^{D/2} \int \frac{d^D \cuts_1 \cdots d^D \cuts_{\ell_b}}{U(t,\theta,\cuts)^{n-f+D\ell_f/2}}. 
\end{equation}

This now can be computed as in the original case we reviewed in \S \ref{FtrickSec} above. We use the change of variables $\cutv=\cuts + \frac{1}{2} M_b(t)^{-1} N_b(\theta)^\tau$. We then have
\begin{equation}\label{svchange}
 \cutv^\tau A_b(t) \cutv = \cuts^\tau A_b(t) \cuts + \frac{1}{2} \cuts^\tau B_b(t,\theta)^\tau + \frac{1}{2} B_b(t,\theta) \cuts + \frac{1}{4} B_b(t,\theta) A_b(t)^{-1} B_b(t,\theta)^\tau,
\end{equation}
where $A_b(t)^\tau=A_b(t)$ and $(B_b(t,\theta)\cuts)^\tau = B_b(t,\theta) \cuts$.

We then rewrite \eqref{intspart} in the form
\begin{equation}\label{intvpart}
\det(M_f(t))^{D/2} \int \frac{d^D \cutv_1 \cdots d^D \cutv_{\ell_b}}{(T +C - \frac{1}{4} B_b A_b^{-1} B_b^\tau + \cutv^\tau A_b \cutv)^{n-f+D\ell_f/2}}. 
\end{equation}

Set  then 
\begin{equation}\label{tildeT}
\tilde T(t,\theta)= T(t,\theta)+ C(t,\theta) - \frac{1}{4} B_b(t,\theta) A_b^{-1}(t) B_b(t,\theta)^\tau,
\end{equation}
so that we write the above as
$$ \frac{\det(M_f(t))^{D/2}}{\tilde T(t,\theta)^{n-f+D\ell_f/2}} 
 \int  \frac{d^D \cutv_1 \cdots d^D \cutv_{\ell_b}}{(1+\cutv^\tau X_b(t,\theta) \cutv)^{n-f+D\ell_f/2}}, $$
with
$$ X_b(t,\theta)= \tilde T (t,\theta)^{-1} A_b(t) . $$
Then, up to a multiplicative constant, the integral gives
\begin{equation}\label{prodform}
 \tilde T^{-n+f-\frac{D\ell_f}{2}+\frac{D\ell_b}{2}} \frac{\det(M_f(t))^{D/2}}{\det(A_b(t))^{D/2}}. 
\end{equation}

\medskip

Consider first the term 
$$ \frac{\det(M_f(t))^{D/2}}{\det(A_b(t))^{D/2}} $$
in \eqref{prodform} above. This can be identified with a Berezinian. In fact, we have
\begin{equation}\label{detBer}
 \frac{\det(M_f(t))^{D/2}}{\det(M_b(t) - \frac{1}{4} M_{fb}(t)
M_f(t)^{-1} M_{fb}(t))^{D/2}} ={\rm Ber}(\cM(t))^{-D/2},  
\end{equation}
where 
\begin{equation}\label{cMt}
 \cM(t)=\left(\begin{array}{cc} M_b(t) & \frac{1}{2} M_{fb}(t) \\
\frac{1}{2} M_{bf}(t) & M_f(t) \end{array}\right). 
\end{equation}

\medskip

We now look more closely at the remaining term $ \tilde
T^{-n+f-\frac{D\ell_f}{2}+\frac{D\ell_b}{2}} $ in \eqref{prodform}. We
know from \eqref{tildeT}, \eqref{AbtBbttheta}, and
\eqref{shortnotation} that we can write $\tilde T(t,\theta)$ in the
form 
\begin{equation}\label{tildeT2}
\tilde T(t,\theta)= \sum_i u_i^2 t_i  + \sum_{j} \cutu_i \theta_{2j-1}
\theta_{2j}  + \sum_{i<j} C_{i j}(t) \theta_{2i-1}\theta_{2i}
\theta_{2j-1} \theta_{2j}, 
\end{equation}
where the first sum is over all edges and the other two sums are over
fermionic edges. We set $\lambda_i = \theta_{2i-1}\theta_{2i}$.  Using
a change of variables $\tilde \lambda_i = \lambda_i + \frac{1}{2} C
\cutu$, we rewrite the above as 
$$ \tilde T(t,\theta)= \sum_i u_i^2 t_i - \frac{1}{4} \cutu^\tau C \cutu
+ \sum_{i<j} C_{ij} \eta_{2i-1}\eta_{2i}\eta_{2j-1}\eta_{2j}, $$
with $\tilde\lambda_i=\eta_{2i-1}\eta_{2i}$.  We denote by
$$ \hat T(t)=\sum_i u_i^2 t_i - \frac{1}{4} \cutu^\tau C \cutu $$
and we write
$$ \tilde T^{-\alpha}= \hat T^{-\alpha} \sum_{k=0}^\infty
{{-\alpha}\choose{k}} \left(\frac{\frac{1}{2}\tilde\lambda^\tau C
\tilde\lambda}{\hat T}\right)^k $$ 
where we use the notation $\frac{1}{2}\tilde\lambda^\tau C \tilde
\lambda =\sum_{i<j} C_{ij} \eta_{2i-1}\eta_{2i}\eta_{2j-1}\eta_{2j}$. 

\medskip

Thus, we can write the Feynman integral in the form
$$ \int \frac{\cutq_1 \cdots \cutq_f}{q_1 \cdots q_n} d^D s_1 \cdots
d^D s_{\ell_b} \, d^D\sigma_1 \cdots d^D \sigma_{\ell_f} = $$
\begin{equation}\label{intLambdaM} 
 \kappa \int_{\Sigma_{n|2f}} \frac{\Lambda(t) \eta_1\cdots \eta_{2f}}
{\hat T(t)^{n-\frac{f}{2}+\frac{D}{2}(\ell_f - \ell_b)} {\rm Ber}(\cM(t))^{D/2}   }
dt_1\cdots dt_n \, d\eta_1 \cdots d\eta_{2f}, \end{equation}
where $\Lambda(t)$ is $\hat T^{f/2}$ times the coefficient of
$\eta_1\cdots \eta_{2f}$ in the expansion 
$$ \sum_{k=0}^\infty {{-\alpha}\choose{k}}
\left(\frac{\frac{1}{2}\tilde\lambda^\tau C \tilde\lambda}{\hat
T}\right)^k. $$ 
More explicitly, this term is of the form
$$ \Lambda(t)= \sum C_{i_1 i_2}(t) \cdots C_{i_{f-1} i_f}(t), $$
over indices $i_a$ with $i_{2a-1}<i_{2a}$ and for $k=f/2$. The
multiplicative constant in front of the integral on the right hand
side above is given by 
$$ \kappa =  {{-n+f-\frac{D}{2}(\ell_f-\ell_b)}\choose{f/2}}. $$

\bigskip

We then obtain the following result.

\begin{thm}\label{BerezinFeynman}
Suppose given a graph $\Gamma$ with $n$ edges, of which $f$ fermionic and $b=n-f$ bosonic. 
Assume that there exists a choice of a basis for $H_1(\Gamma)$ satisfying the condition
\begin{equation}\label{combrelFB}
n-\frac{f}{2} + \frac{D}{2}(\ell_f -\ell_b) =0.
\end{equation}
Then the following identity holds:
\begin{equation}\label{intqintBer}
\int \frac{\cutq_1 \cdots \cutq_f}{q_1 \cdots q_n} d^D s_1 \cdots d^D s_{\ell_b} \, d^D\sigma_1 \cdots d^D \sigma_{\ell_f} = \int_{\Sigma_n} \frac{\Lambda(t)}{{\rm Ber}(\cM(t))^{D/2}} dt_1\cdots dt_n .
\end{equation}
\end{thm}

\proof
This follows directly from \eqref{intLambdaM}, after imposing $n-\frac{f}{2} + \frac{D}{2}(\ell_f -\ell_b) =0$ and performing the Grassmann integration of the resulting term
\begin{equation}\label{intSigman2fS}
 \int_{\Sigma_{n|2f}} \frac{\Lambda(t)\eta_1\cdots\eta_{2f}}{{\rm Ber}(\cM(t))^{D/2}} dt_1\cdots dt_n \, d\eta_1 \cdots d\eta_{2f} . 
\end{equation}
\endproof

\subsection{Graph supermanifolds}

The result of the previous section shows that we have an analog of the period integral 
$$ \int_{\Sigma_n} \frac{dt_1 \cdots dt_n}{\det(M_\Gamma(t))^{D/2}} $$
given by the similar expression
\begin{equation}\label{intSigmanS}
 \int_{\Sigma_n} \frac{\Lambda(t)}
{{\rm Ber}(\cM(t))^{D/2}} dt_1 \cdots dt_n . 
\end{equation}
Again we see that, in this case, divergences arise from the intersections between the domain of integration given by the simplex $\Sigma_n$ and the subvariety of $\P^{n-1}$ defined by the 
solutions of the equation
\begin{equation}\label{BLzero}
 \frac{{\rm Ber}(\cM(t))^{D/2}}{\Lambda(t)} =0 . 
\end{equation} 

\begin{lem}\label{DivisorML}
For generic graphs, the set of zeros of \eqref{BLzero} defines a hypersurface in $\P^n$, hence a divisor in $\P^{n-1|2f}$ of dimension $(n-2|2f)$. The support of this divisor is the same as that of the principal divisor defined by ${\rm Ber}(\cM(t))$.
\end{lem}

\proof The generic condition on graphs is imposed to avoid the cases with $M_f(t)\equiv 0$. Thus, suppose given a pair $(\Gamma,B)$ that is generic, in the sense that $M_f(t)$ is not identically zero.
The equation \eqref{BLzero} is satisfied by solutions of $$\det(M_b(t)-\frac{1}{4} M_{bf}(t) M_f(t)^{-1} M_{fb}(t)) =0$$ and by poles of $\Lambda(t)$. Using the formulae \eqref{AbtBbttheta} and \eqref{shortnotation} we see that the denominator of $\Lambda(t)$ is given by powers of $\det(M_f(t))$ and $\det (A_b(t))= \det(M_b(t)-\frac{1}{4} M_{bf}(t) M_f(t)^{-1} M_{fb}(t))$. Thus, the set of solutions of \eqref{BLzero} is the union of zeros and poles of ${\rm Ber}(\cM(t))$. The multiplicities are given by the powers of these determinants that appear in $\Lambda(t) {\rm Ber}(\cM(t))^{-D/2}$.
\endproof 

\begin{defn}\label{superGammaDef}
Let $\Gamma$ be a graph with bosonic and fermionic edges and $B$ a choice of a basis of $H_1(\Gamma)$. We denote by $\cX_{(\Gamma,B)} \subset \P^{n-1|2f}$ the locus of zeros and poles of
${\rm Ber}(\cM(t)) =0$. We refer to $\cX_{(\Gamma,B)}$ as the {\em graph supermanifold}.
\end{defn}

In the degenerate cases of graphs such that $M_f(t)\equiv 0$,  we simply set $\cX_{(\Gamma,B)}=\P^{n-1|2f}$. Examples of this sort are provided by data $(\Gamma,B)$ such that there is only one loop in $B$ containing fermionic edges.  
Other special cases arise when we consider graphs with only bosonic or only fermionic edges. In the first case, we go back to the original calculation without Grassmann variables and we therefore simply recover $\cX_{(\Gamma,B)}= X_\Gamma= \{ t: \det(M_b(t))=0\} \subset \P^{n-1|0}$.
In the case with only fermionic edges, we have $\det(M_b(t)-\frac{1}{4} M_{bf}(t) M_f(t)^{-1} M_{fb}(t))\equiv 0$ since both $M_b(t)$ and $M_{bf}(t)$ are identically zero. It is then natural to simply assume that, in such cases, the graph supermanifold is simply given by $\cX_{(\Gamma,B)}=\P^{f-1|2f}$.

\subsection{Examples from Feynman graphs}

\begin{center}
\begin{figure}
\includegraphics[scale=0.5]{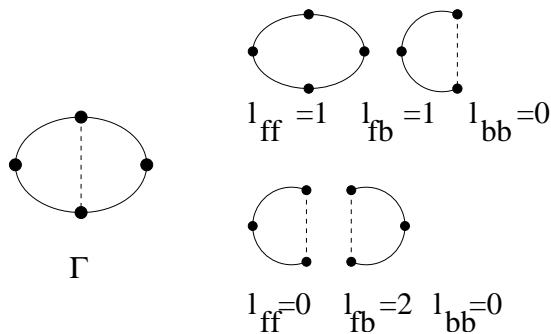}
\caption{Choices of a basis for $H_1(\Gamma)$.
\label{GraphH1Basis}}
\end{figure}
\end{center}

We still need to check that the condition \eqref{combrelFB} we imposed on the graph is satisfied by some classes of interesting graphs. First of all, notice that the condition does not depend on the graph alone, but on the choice of a basis for $H_1(\Gamma)$. The same graph can admit choices for which 
\eqref{combrelFB} is satisfied and others for which it fails to hold. 
For example, consider the graph illustrated in Figure \ref{GraphH1Basis}, for a theory in dimension $D=6$, where we denoted bosonic edges by the dotted line and fermionic ones by the full line. There exists a choice of a basis of $H_1(\Gamma)$ for which  \eqref{combrelFB} is satisfied, as the first choice in the figure shows, while not all choices satisfy this condition, as one can see in the second case.

\begin{center}
\begin{figure}
\includegraphics[scale=0.5]{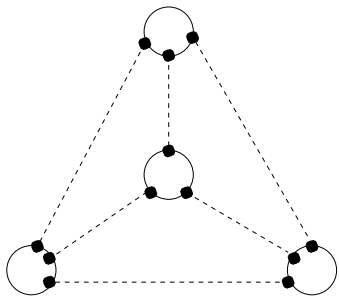}
\caption{A graph with a basis of $H_1(\Gamma)$ satisfying \eqref{combrelFB}.
\label{Graph2rel}}
\end{figure}
\end{center}

One can see easily that one can construct many examples of graphs that admit a basis of $H_1(\Gamma)$ satisfying \eqref{combrelFB}. For instance, the graph in Figure \ref{Graph2rel} is a slightly more complicated example in $D=6$ of a graph satisfying the condition. Again we used dotted lines for the bosonic edges and full lines for the fermionic ones.

\begin{center}
\begin{figure}
\includegraphics[scale=0.5]{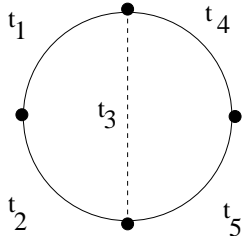}
\caption{Edge variables.
\label{LabelEdges}}
\end{figure}
\end{center}

Let us consider again the example of the very simple graph of Figure \ref{GraphH1Basis}, with the first choice of the basis $B$ for $H_1(\Gamma)$. This has two generators, one of them a loop made of fermionic edges and the second a loop containing both fermionic and bosonic edges. Let us assign the ordinary variables $t_i$ with $i=1,\ldots,5$ to the edges as in Figure \ref{LabelEdges}. We then have
$$ M_b(t)=t_1+t_2+t_3 $$
since only the second loop in the basis contains bosonic edges, while we have
$$ M_{bf}(t)= (t_1+t_2, t_1+t_2+t_3)=t_1 (1,1)+t_2 (1,1) +t_3 (0,1) + t_4(0,0) +t_5 (0,0) $$
and
$$ M_f(t)=\left(\begin{array}{cc} 0 & t_1+t_2 \\ -(t_1+t_2) & 0 \end{array}\right). $$
Thus, we obtain in this case
$$ M_{bf}(t) M_f(t)^{-1} M_{fb}(t)= (t_1+t_2, t_1+t_2+t_3) \left(\begin{array}{cc} 0 & \frac{-1}{t_1+t_2} \\ \frac{1}{t_1+t_2} & 0 \end{array}\right) \left(\begin{array}{c} t_1+t_2 \\ t_1+t_2+t_3 \end{array}\right) $$ 
$$ = (t_1+t_2, t_1+t_2+t_3) \left(\begin{array}{c} \frac{-( t_1+t_2+t_3)}{t_1+t_2} \\ 1
 \end{array}\right)=-( t_1+t_2+t_3)+ t_1+t_2+t_3 \equiv 0.
$$
Thus, in this particular example we have $M_{bf}(t) M_f(t)^{-1} M_{fb}(t)\equiv 0$ for all $t=(t_1,\ldots, t_5)$, so that ${\rm Ber}(\cM(t))=\det(M_b(t)) \det(M_f(t))^{-1}=(t_1+t_2+t_3)/(t_1+t_2)^2$ and the locus of zeros and poles $\cX_{(\Gamma,B)}\subset \P^{5|8}$ is the union of $t_1+t_2+t_3=0$ and $t_1+t_2=0$ in $\P^5$ (the latter counted with multiplicity two), with the restriction of the sheaf from $\P^{5|8}$.

\subsection{The universality property}

Lemma \ref{superBB} below shows to what extent the ``universality property" 
of graph hypersurfaces, \ie the fact that they generate the Grothendieck group of varieties, continues to hold when passing to supermanifolds.

\begin{lem}\label{superBB}
Let $\cR$ be the subring of the Grothendieck ring $K_0(\cS\cV_\C)$ of supermanifolds spanned by the $[\cX_{(\Gamma,B)}]$, for $\cX_{(\Gamma,B)}$ the graph supermanifolds defined by the divisor of zeros and poles of the Berezinian ${\rm Ber}(\cM(t))$, with $B$ a choice of a basis of $H_1(\Gamma)$. Then  $$ \cR =K_0(\cV_\C)[T^2]\subset  K_0(\cS\cV_\C), $$
where $T=[\A^{0|1}]$.
\end{lem}

\proof By Corollary \ref{ringSGr} and the universality result of \cite{BeBro}, it suffices to prove that the subring of $K_0(\cS\cV_\C)$ generated by the $[\cX_{(\Gamma,B)}]$ contains the classes of the ordinary graph hypersurfaces in $K_0(\cV_\C)$ and the class $[\A^{0|2}]$. 

\begin{center}
\begin{figure}
\includegraphics[scale=0.5]{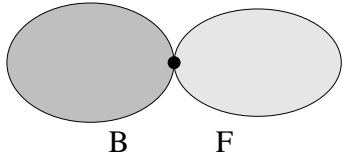}
\caption{Graphs with $\ell_{bf}=0$.
\label{BFgraphs}}
\end{figure}
\end{center}

To show that $\cR$ contains the ordinary graph hypersurfaces, consider the special class of graphs that are of the form schematically illustrated in Figure \ref{BFgraphs}. These are unions of two graphs, one only with bosonic edges and one only with fermionic edges, with a single vertex in common. Notice that in actual physical theories the combinatorics of graphs with only fermionic edges is severely restricted (see \cite{Ramond},  \S 5.3)
depending on the dimension $D$ in which the theory is considered. However, for the purpose of this universality result, we allow arbitrary $D$ and corresponding graphs, just as in the result of \cite{BeBro} one does not restrict to the Feynman graphs of any particular theory. 

The graphs of Figure \ref{BFgraphs} provide examples of graphs with bases of $H_1(\Gamma)$ containing loops with only fermionic or only bosonic legs, \ie with $\ell_{bf}=0$, $\ell_f=\ell_{ff}$ and $\ell_b=\ell_{bb}$. This implies that, for all these graphs $\Gamma=\Gamma_B \cup_v \Gamma_F$ with the corresponding bases of $H_1$, one has $M_{bf}(t)\equiv 0$, since for each edge variable $t_i$ one of the two factors $\eta_{i r_b} \eta_{i r_f}$ is zero. Thus, for this class of examples we have ${\rm Ber}(\cM(t))=\det(M_b(t))/\det(M_f(t))$. Moreover, we see that
for these examples $\det(M_b(t))=\Psi_{\Gamma_b}(t)$ is the usual graph polynomial of the graph $\Gamma_B$ with only bosonic edges. Since such $\Gamma_B$ can be any arbitrary ordinary graph, we see that the locus of zeros alone, and just for this special subset of the possible graphs, already suffices to generate the full $K_0(\cV_\C)$ since it gives all the graph varieties $[X_{\Gamma_B}]$. 

To show then that the subring $\cR$ contains the classes $[\A^{0|2f}]$, for all $f$, first notice that the classes $[\P^n][\A^{0|2f}]=[pt][\A^{0|2f}]+ [\A^{1|0}][\A^{0|2f}]+\cdots+[\A^{n|0}][\A^{0|2f}]$ belong to $\cR$, for all $n$ and $f$. These are supplied, for instance, by the graphs with a single loop containing fermionic edges, as observed above. This implies that elements of the form $[\A^{n|0}][\A^{0|2f}]=[\P^n][\A^{0|2f}]-[\P^{n-1}][\A^{0|2f}]$ belong to $\cR$. In particular the graph consisting of a single fermionic edge closed in a loop gives $[\A^{0|2f}]$ in $\cR$.  
\endproof

Notice that in \cite{BeBro}, in order to prove that the correspoding graph hypersurfaces generate $K_0(\cV_\C)$,  one considers all graphs and not only the log divergent ones with $n=D\ell/2$, even though only for the log divergent ones the period has the physical interpretation as Feynman integral. Similarly, here, in Lemma \ref{superBB}, we consider all $(\Gamma,B)$ and not just those satisfying the condition
\eqref{combrelFB}.

The fact that we only find classes of the even dimensional superplanes $[\A^{0|2f}]$ in $\cR$ instead of all the possible classes $[\A^{0|f}]$ is a consequence of the {\em fermion doubling} used in Lemma \ref{SigmanfInt} in the representation of the Feynman integral in terms of an ordinary and a fermionic integration.

\section{Supermanifolds and mirrors}\label{MSsect}

We discuss here some points of contact between the construction we outlined in
this paper and the supermanifolds and periods that appear in the theory of
mirror symmetry. 

Supermanifolds arise in the theory of mirror symmetry (see for
instance \cite{Se}, \cite{AgVa}, \cite{KuPo}) in order to describe mirrors of
rigid Calabi--Yau manifolds, where the lack of moduli of complex structures
prevents the existence of K\"ahler moduli on the mirror. The mirror still
exists, not as a conventional K\"ahler manifold, but as a supermanifold
embedded in a (weighted) super-projective space. 

For instance, in the construction given in \cite{Se}, one considers the hypersurface in 
(weighted) projective space given by the vanishing of a superpotential $X=\{ W=0 \}
\subset \P^n$. The local ring of the hypersurface $X$ is given by polynomials in the
coordinates modulo the Jacobian ideal $\cR_X =\C[x_i]/dW(x_i)$. To ensure the vanishing 
of the first Chern class, one corrects the superpotential $W$ by additional quadratic
terms in either bosonic or fermionic variables, so that the condition $W=0$ defines
a supermanifold embedded in a (weighted) super-projective space, instead of an ordinary 
hypersurface in projective space.

In the ordinary case, one obtains the primitive part of the middle cohomology 
$H^{n-1}_0(X)$ and its Hodge decomposition via the Poincar\'e residue 
\begin{equation}\label{Res}
Res(\omega)=\int_C \omega,
\end{equation}
with $C$ a 1-cycle encircling the hypersurface $X$, applied to forms
of the form
\begin{equation}\label{ResW}
\omega(P)= \frac{P(x_0,\ldots,x_n)\Omega}{W^k},
\end{equation}
with $\Omega=\sum_{i=0}^n (-1)^i \lambda_i x_i dx_0 \cdots \widehat{dx_i}\cdots dx_n$, 
as in \eqref{Omegasum} with $\lambda_i$ the weights in the case of weighted projective
spaces, and with $P \in \cR_X$ satisfying $k\deg(W)=\deg(P)+\sum_i \lambda_i$.

In the supermanifold case, one replaces the calulation of the Hodge structure on
the mirror done using the technique described above, by a supergeometry analog,
where the forms \eqref{ResW} are replaced by forms 
\begin{equation}\label{sResW}
\frac{P(x_0,\ldots,x_n) d\theta_1\cdots d\theta_{2m} \Omega}{W^k},
\end{equation}
where here the superpotential $W$ is modified by the presence of an additional quadratic term
in the fermionic variables $\theta_1\theta_2+\cdots\theta_{2m-1}\theta_{2m}$. 

\smallskip

In comparison to the setting discussed in this paper, notice that the procedure of replacing the 
potential $W$ by $W'=W+\theta_1\theta_2+\cdots\theta_{2m-1}\theta_{2m}$, with the additional
fermionic integration, is very similar to the first step in our derivation where we replaced the
original expression $T=t_1q_1+\cdots t_n q_n$ by the modified one $T+\frac{1}{2} \theta^\tau \cQ \theta$ 
with $\frac{1}{2} \theta^\tau \cQ \theta=\cutq_1 \theta_1\theta_2+\cdots +\cutq_f \theta_{2f-1}\theta_{2f}$.
Thus, replacing the ordinary integration $\int T^{-n}(t) dt$ by the integration
$\int (T(t)+\frac{1}{2} \theta^\tau \cQ \theta)^{-n+f} dt d\theta$
is an analog of replacing the integral $\int W^{-k} dt$ with the integral $\int (W+\theta_1\theta_2+\cdots\theta_{2m-1}\theta_{2m})^{-k} dt d\theta$ used in the mirror symmetry context. However, there seems
to be no analog, in that setting, for the type of periods of the form \eqref{intSigmanS} that we obtain here
and for the corresponding type of supermanifolds defined by divisors of Berezinians considered here.

\end{document}